\documentstyle{article}

\newcount\Mac  \Mac=0  
\newcommand{\ifMac}[2]{\ifnum\Mac=1 #1 \else #2 \fi}

\def\be{\begin{equation}}
\def\ee{\end{equation}}
\def\bea{\begin{eqnarray}}
\def\eea{\end{eqnarray}}

\begin{document}
\title{\huge\bf Unified Theories}
\author{\bf Riccardo Barbieri\\ \em Physics Department, University of Pisa\\
 \em    and INFN, Sez.\ di Pisa, I-56126 Pisa, Italy}

\maketitle
\begin{abstract}
{The present status of Unified Theories is summarized with special emphasis
on their possible experimental tests. Outline: i) Unification
of couplings; ii) Where can a positive signal come from? iii) HERA anomaly
and Unification; iv) Recent progress in model building; v) Flavour and
Unification.}
\end{abstract}
  
\section{Introduction and unification of couplings} \label{sec:unification}
In the description of the fundamental
interactions among elementary particles, symmetries play a central role, more
important than  particles themselves. The diversity of particles and/or of
their
interactions can actually be often understood,
in a way or another,
as a manifestation of the underlying basic symmetries. As we know, three
kinds of symmetries are or may be of relevance in these respects:
space-time symmetries,
intra-family (vertical) symmetries, inter-family (horizontal or flavour)
symmetries.
There is clear evidence for space-time symmetries, which might include
supersymmetry, as there is compelling evidence for the vertical SU(3,2,1)
gauge symmetry.
On the contrary, the role and the nature of flavour symmetries is still
controvertial.

Unified theories can mostly be viewed as attempts to enlarge the role of these
symmetries in the description of elementary particle physics. There is, in
fact,
circumstantial evidence in favour of an enlargement of the vertical SU(3,2,1)
gauge
symmetry to a more unified group, like SU(5) or bigger. Such evidence is
both of algebric and of empirical nature. The quantum numbers of the quarks
and leptons of one generation fit into simple representations of the unified
group, as observed more than twenty years ago~\cite{ggps}.
Furthermore, the more recent
precise measurements of the weak mixing angle $sin^2 \theta_W$ and of the
strong coupling constant $\alpha_S (M_Z)$ indicate their meeting with
the electromagnetic coupling constant $\alpha (M_Z)$ at a large energy scale
$M_G$~\cite{gqw}, provided the "low energy" spectrum includes
the new particles implied by supersymmetry~\cite{drw,dg1}: to make
a larger vertical symmetry working calls for an enlargement of the space-time
symmetry, including supersymmetry,
 as also demanded by the need to stabilize the huge
hierarchy between the weak and the unification scales. This is illustrated in
Fig. 1, which shows the prediction for $\alpha_S (M_Z)$ as function of an
appropriately defined "mean" supersymmetric mass~\cite{lang}, $T_{susy}$,
ranging
between
$10$ and $100$ $GeV$ in the "natural" region of parameter space.
The band of the "SUSY GUT" prediction
corresponds to what I think is a reasonable definition of the
theoretical uncertainties associated
with physics occurring at the high energy scale, like heavy
thresholds~\cite{GR} and/or
Planck scale  non renormalizable effects~\cite{NA}\footnote{Not displayed
in Fig. 1 is a
low energy threshold effect due to light superpartners which may, but need
not, increase the
prediction for $\alpha_S (M_Z)$ by about $5 \%$~\cite{fg}}. The
agreement of such
prediction with the current world average,
$\alpha_S (M_Z)=0.119 \pm 0.005$~\cite{cat}, is remarkable. On the other hand,
this figure makes
clear that the unification of couplings does not set an upper bound on the
scale of
supersymmetric  particle masses relevant to their foreseable experimental
search. On the
contrary, it determines with significant precision the value of the
unification scale.

\begin{figure}[t]\setlength{\unitlength}{1cm}
\begin{center}\begin{picture}(15.5,9)
\ifMac
{\put(0,0){\special{picture GUT}}}
{\put(0,1){\includegraphics{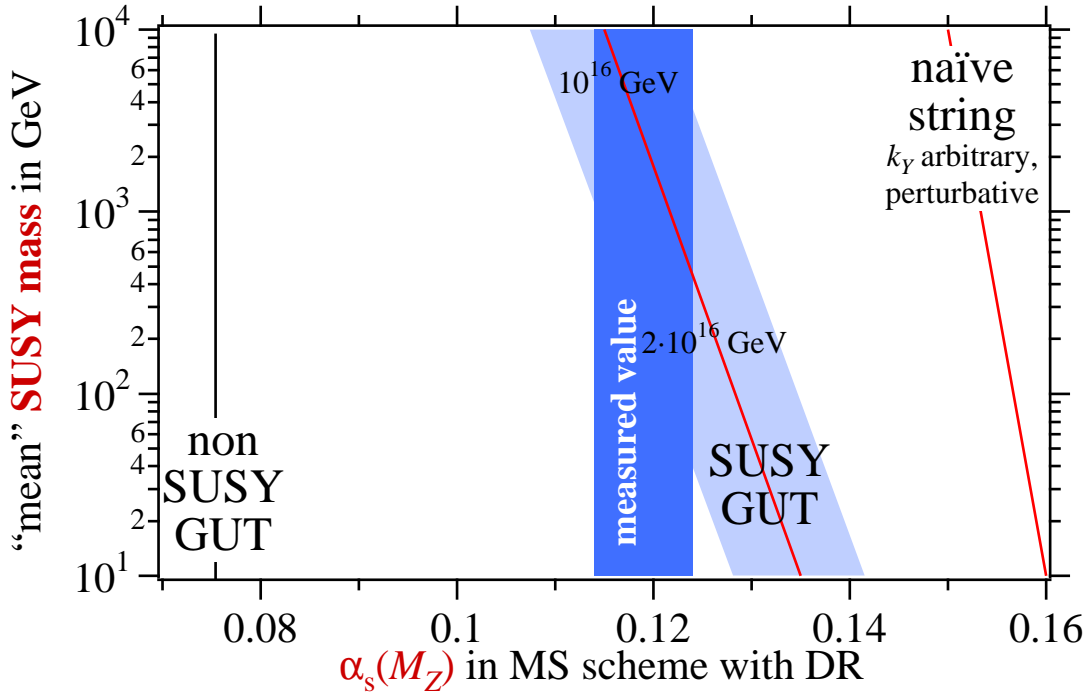}}}
\end{picture}
\caption[fig1]{\em $\alpha_{\rm s}$-prediction in supersymmetric GUT with
minimal particle content.\label{fig:eRN}}
\end{center}
\end{figure}

Fig. 1 also shows the central value of the "string theory prediction", defined
by requiring, with the same spectrum as in the "SUSY GUT" case, the meeting
of the
three couplings at the string scale $M_{st}=4\cdot 10^{17}GeV$~\cite{kap}
with an arbitrary
Kac Moody coefficient for the hypercharge U(1) factor. The discrepancy between
$M_{st}$ and $M_G$, as determined by Fig. 1 itself, is the source of the
problem.
By no means, however, this
must be viewed as a conflict between field theory and string theory. String
theory, which actually forces unification of the gauge couplings also with the
gravitational coupling, can still require $\alpha_S (M_Z)$ to be in
agreement with
the observed value
in several ways. At least
three such ways have been
put forward: i) a stage of Grand Unification, in the usual field theory
sense, below the
string scale~\cite{NA8}\footnote{Provided the spectrum below $M_G$ is the
one of the
Minimal Supersymmetric Standard Model};  ii) a modification of space-time
below the GUT scale
which influences the  energy dependence of the gravitational constant but
not of the gauge
couplings~\cite{RO3}; iii) the addition of extra matter multiplets at
intermediate energies
or close to the GUT scale~\cite{by}. In particular only i) requires an
intermediate stage of
field-theoretical Grand Unification, which, therefore, may or may not be
necessary. For model
building it is important to have a view on the relative plausibility of
these two
alternatives. Of help, to this purpose, it would be to know better which
are the
consequences, at the large scale, of the non-GUT picture, as derived from
string theory. Can
we distinguish the two alternatives in a phenomenological way other than
discussing the
unification of couplings itself? Hereafter, when important, by Unification
I will mean Grand
Unification.

\section{Where can a positive signal come from?} \label{sec:signals}

It is both logical and useful, at this stage, to briefly recall the ways to
find
an experimental signal of supersymmetry and Unification. They are summarized
in Table 1, togheter with a qualification of their significance. All boxes of
this Table require, as always, an element of judgement. Let me briefly comment
on each of them, not before having noticed that an entry on neutrino masses
might have been added to the list. Neutrino masses are indeed generally
expected in a Unified Theory. The reason for not having such an entry
in Table 1 is twofold: neutrino masses are not a specific
prediction of Unified Theories only, nor there is a value for
neutrino masses that Unified Theories clearly prefer, at least in my view.
Having said that,
I will have to defend the insertion of "Non Standard FCNC" in Table 1.
The last part of the talk is devoted to this issue.

\begin{table}
\begin{center}
\begin{tabular}{|p{1in}||p{2cm}p{2cm}p{2cm}|}\hline
Find & Sufficient & Strong Indication & Necessary\\ \hline \hline
Superpartners below 1 TeV & $\surd$&&$\surd$\\ \hline
Proton decay &$\surd$&&\\ \hline
WIMP Dark Matter &&$\surd$&\\ \hline
A light Higgs &&$\surd$&$\surd$\\ \hline
Non Standard FCNC &&$\surd$&\\
\hline
\end{tabular}
\caption{\em Summary of possible signals for supersymmetry and Unification.}
\end{center}
\end{table}

\begin{figure}[t]\setlength{\unitlength}{1cm}
\begin{center}\begin{picture}(15.5,9)
\ifMac
{\put(0,0){\special{picture eRN}}}
{\put(0,1){\includegraphics{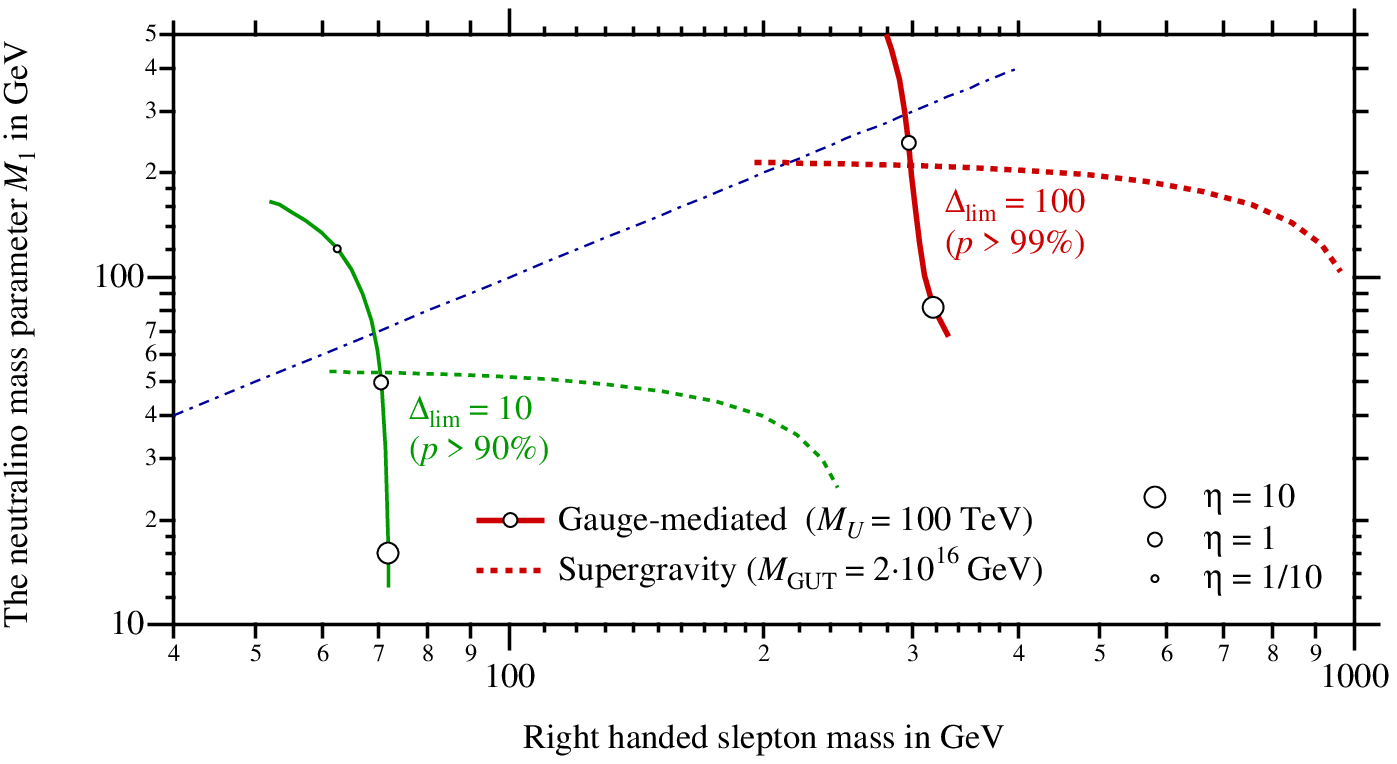}}}
\end{picture}
\caption[fig1]{\em Naturalness bounds on the masses of the lightest 
neutralino and of the lightest slepton in gauge mediation and
in supergravity models\label{fig:eRN}}
\end{center}
\end{figure}

Although somewhat model dependent, "naturalness" bounds hold on the
supersymmetric
particle masses, requiring some of them to be below 1 TeV and, in
some cases, well below that limit~\cite{nat}. The relationship between
the Z-mass, or the Fermi scale, and the
masses of the superpartners allows to define a
"natural" region of the parameter space. In supergravity models, the
exploration of $90 \%$ of this region (or allowing up to a 10
$\%$ fine-tuning among the various parameters) gives upper limits on the
masses of the gluino, the lightest stop, the
lightest chargino and the lightest neutralino of about
$350, 300, 90, 50$  $GeV$ respectively. This region is therefore being
significantly
explored by present (so far negative) searches~\cite{dio}. In general,
these limits scale as
the square-root of the ratio  of the allowed amount of fine tuning. As
such, even a
$99 \%$ probability bound still requires all these particles to be lighter than
about
$1TeV$.  Bounds weaker by about one order of magnitude apply to the
sfermions of the
first two generations, because of their small Yukawa
couplings to the Higgs~\cite{dgi}.

As discussed later on, "gauge-mediation" models~\cite{gm,gmn} are
considered as  alternative to "supergravity" models~\cite{bfs} in the
description of
supersymmetry breaking. Since the spectrum of superpartners changes
in the two cases, so do the naturalness bounds, as illustrated in Fig. 2,
taken from Ref.
~\cite{cs} (See also Ref. ~\cite{grr}). There the bounds ( $p>90 \%$ or $p>99
\%$) are
compared in the two cases for the lightest neutralino and for the
right-handed stau,
taking into account the correlation in the respective parameter spaces.
All this substanciates the statement that finding superpartners below $1TeV$
is a necessity in Unified Theories. Conversely, discovering superpartners in
this energy range would prove supersymmetry as a relevant symmetry in nature
but would still only be an indirect sign of Unification, unless detailed
measurements of the superpartner spectrum are made.

The detection of proton decay appears as second entry in Table 1, in fact under
the column "Sufficient". Unfortunately, this search cannot be also qualified
as "Necessary". The only firm prediction of supersymmetric Unification is
the rate for the $e^+ \pi^0$ mode, mediated by vector boson exchanges,
\begin{equation}
\tau(p\rightarrow e^+ \pi^0)=1\cdot 10^{35 \pm1}(M_G/10^{16}GeV)^4 years
\label{tau}
\end{equation}
to be compared with the present limit on this mode of $6 \cdot10^{32}$
years~\cite{PDG}
and with the expected sensitivity of the Superkamiokande detector, after 10
years
of running, of about $10^{34}$ years. More favourable modes could be
$p\rightarrow K^+ \nu_\mu$, $p\rightarrow K^0 \mu^+$, $n\rightarrow K^0
\nu_\mu$,
since they might occur at higher, but also more
uncertain rates than (\ref{tau})~\cite{gr,lr}.

Several detectors are active in the world searching for dark matter
in our galaxy in the form of Weakly Interacting Massive Particles. A positive
signal would be a significant indication in favour of supergravity
type models. In a relevant portion of their natural
parameter space, these models require the lightest neutralino to be
a cold dark matter particle, as the relic abundance
of neutralinos from the big bang
can be safely calculated~\cite{hg,bb}. Within an uncertainty factor of
about 5, it is
also possible to estimate the expected event rate, per kg, per day, in
a given WIMP dark matter detector. Current detectors are not yet capable
to compete significantly with direct searches of supersymmetric particles,
except in some corners of parameter space~\cite{bb}.
They will, however, when sensitivities considered achievable with present
technology (e.g. 0.01 events/kg/day in a Ge or in a NaI detector) are
reached.

The discovery of a light Higgs should also be viewed as a very strong
indication in favour of supersymmetry and, indirectly, of supersymmetric
Unification, even without knowing if its properties deviate from those
expected for the Standard Model Higgs. Supersymmetry predicts a full set
of Higgs particles, among which one at least should be relatively light:
less than about $120-130$ $ GeV$~\cite{lH} in the model with minimal low
energy
particle content and
than about
$150-160$ $GeV$~\cite{lHg} in a generic model.

Finally the last entry in Table 1 is discussed with special attention below.

\section{HERA anomaly and Unification} \label{sec:HERA}

An anomaly of the high-$Q^2$ events in $e^+p$ scattering has been presented
at this Conference~\cite{HERA} and it has been discussed as a possible
evidence for
new physics~\cite{HERA,gual}. Such an anomaly could be due, although in
any case not without difficulties, to some new contact
interaction or to the production of an s-channel resonance~\cite{gual}.
Since Unification is a
leading candidate for extending the Standard Model, it is natural to ask if and
how this anomaly, supposedly confirmed by further necessary data, could be
described
in a Unified picture.

Two different contact interactions, or
a combination thereof, invariant under $SU(3,2,1)$, might perhaps be responsible
of the HERA-anomaly and not be in conflict with any other experiment so
far. They
are composed by the following two chains of 4-fermion interactions~\cite{gual,ddx} ($L$ and
$Q$ are lepton
and quark doublets, of given chirality)
\begin{equation}
L_L\gamma_{\mu}L_L Q_R\gamma_{\mu}Q_R +
L_R\gamma_{\mu}L_R Q_L\gamma_{\mu}Q_L,
\label{ct1}
\end{equation}
\begin{equation}
L_R\gamma_{\mu}L_R (Q_L\gamma_{\mu}Q_L - Q_R\gamma_{\mu}Q_R),
\label{ct2}
\end{equation}
weighted by an appropriate inverse squared energy scale \footnote{The
right-handed-neutrino in $L_R$ is there only for simplicity of notation}.
In short, I do not know how to make these effective interactions emerge
from any sort of Unified Theory with decent
assumptions about the dynamics or the symmetry
breaking pattern. The need to sufficiently suppress any other
similar interaction not seen in other experiments or in HERA itself is the
major difficulty.

At least in principle, it is easier to accommodate in a
Unified Theory a new particle being
exchanged in the s-channel, in
particular a scalar leptoquark, as the origin of the HERA-anomaly. To
make the game less wild, let me consider the possibility that
this leptoquark is actually a s-quark with a superpotential
coupling of the form
\begin{equation}
\lambda'_{113} L_1 D_1 Q_3,
\label{l113}
\end{equation}
 where
$L_1, D_1, Q_3$ are the left-handed lepton doublet of the first
generation, the down-type quark singlet of the first generation and the
third generation quark doublet respectively. The scalar top in $Q_3$,
with a mass of about $200 GeV$,
would be the leptoquark supposedly produced at HERA with a
 coupling $\lambda'_{113}$ at the few percent level~\cite{gual}.

Even this interpretation is not without "potential" difficulties, however.
One is generic: the presence of other $\lambda'$ couplings, similar in
strenght to  $\lambda'_{113}$ but with different flavour
indices, would induce unobserved FCNC interactions~\cite{hd}. Another
difficulty is more tied to the very concept of unification: why
not to have, togheter with (\ref{l113}), also the couplings ($E$ is the
charged lepton
singlet and $U$ the up-type quark singlet)
\begin{equation}
\lambda_{ijk} L_i L_j E_k + \lambda''_{ijk} D_i D_j U_k,
\label{l}
\end{equation}
as a genuine full "vertical" symmetry would suggest? I have
called these difficulties "potential" because it is possible
to set initial conditions on fully unified couplings, for
example $SU(5)$-invariant ones, which at low energy,
consistently with renormalization rescalings, avoid these
problems~\cite{bbs}. If more data were to confirm the anomaly
and strenghten its leptoquark interpretation, it would
become interesting to see if and how these initial conditions
could be made natural in some sense~\cite{ggr}.

\section{Recent progress in model building} \label{sec: models}

In model building of Unified Theories, two main problems remain open, at
least in the sense
that no possible solution clearly emerges yet over the others:
\begin{itemize}
\item[i)] The supersymmetry breaking problem. One has to
originate the soft supersymmetry breaking
parameters in the scalar potential as well as the so called "$\mu$-term". All
these parameters are characterized by a scale $\Lambda_{SB}$, defined as the
scale at which the corresponding Lagrangian terms cease to appear as local
interactions.

\item[ii)] The
flavour symmetry breaking problem. Here the relevant parameters are
the Yukawa couplings, also depending on a scale  $\Lambda_{FB}$, defined
in an analogous way to $\Lambda_{SB}$.
\end{itemize}
Both these problems are not
particularly new. Some new inputs, however, have influenced the recent
developments
in this area. One is a technical tool: the dynamics of strongly coupled
supersymmetric field theories is better understood, mostly due to the
work of Seiberg and Seiberg and Witten~\cite{sw}, further developing
earlier work in the
eighties~\cite{eit}. The second input is due to a better focus on the
connection
that actually exists between the two problems mentioned above. Such connection
 makes it useful to devide the possible theories in two classes, depending
on the
relation among $\Lambda_{FB}$, $\Lambda_{SB}$ and $M_G$:

1) "supergravity-type" theories, characterized by $\Lambda_{SB}\geq
\min(\Lambda_{FB}, M_G)$;

2) "gauge-mediation-type" theories, for which $\Lambda_{SB}\ll
\min(\Lambda_{FB}, M_G)$.

From a general point of view, the important difference between these two
classes
of theories is that in the first case, unlike the second one, the supersymmetry
breaking parameters necessarily feel also the breaking of flavour, at least
as an
effect of radiative corrections. Such coupling was pointed out long time
ago already in the context of the Minimal Supersymmetric Standard
Model~\cite{nw} and later
realized to be even more important in the GUT case~\cite{hkr}, especially
because of the
heavyness of the top quark~\cite{bh}.

As a result of these new inputs, special efforts have been made in studying
explicit renormalizable field theory models of dynamical supersymmetry
breaking.
This is relevant in the case of "gauge-mediation-type" theories, since
there, by
definition, $\Lambda_{SB}$ is lower than $M_G$, a scale at which renormalizable
field theory should already be effective. As a result, using strongly
coupled theories, several mechanisms for dynamical supersymmetry breaking
have been
designed, too many to be described in detail. The following might at least
be a not
too incomplete list:
\begin{enumerate}
\item Strong coupling models~\cite{eit};
\item Dynamically generated superpotentials~\cite{dgs};
\item Confinement without chiral symmetry breaking~\cite{csb};
\item Quantum modified constraints~\cite{qmc};
\item Product groups~\cite{pg};
\item Plateau of supersymmetry breaking false vacua~\cite{psb}.
\end{enumerate}Are these mechanisms any useful? I think they are, since they make possible the
construction of complete and relatively simple realistic examples of Unified
Theories based on gauge-mediation, putting them on essentially equal
footing than
models based on supersymmetry breaking transmitted by supergravity
couplings. Some
further progress in this direction might still occur.

Is it then a "supergravity-type" or a "gauge-mediation-type" theory which
is realized in nature, if any? Experiments can and must decide, on the
basis of the
following phenomenological differences:
\begin{itemize}
\item[i)] Different spectra of the superpartners, with, e.g, s-quarks and s-leptons
significantly
more separated in mass in the "gauge-mediation" case~\cite{gms}.

\item[ii)] Different behaviour of the "Lightest Supersymmetric Particle". In
"gauge-mediation" the LSP is always the gravitino, to which the lightest among
the usual superpartners, most likely a neutralino or a stau, decays. The wide
variety of possible lifetimes, consequence of a largely unconstrained gravitino
mass, may lead, in turn, to very different experimental signatures:
prompt 2 $\gamma$'s, delayed $\gamma$'s, heavy charged tracks, etc.

\item[iii)] Different flavour physics. Although this is not a theorem, I am convinced
that any sensible Unified Theory based on supergravity will significantly
deviate from the SM
expectations in some flavour physics observable~\cite{bh}, whereas this
need not be the case
in gauge-mediation models.
\end{itemize}
This brings me to the last subject of this talk, on which I wish to expand
my comments a bit.

\section{Flavour and Unification} \label{sec: FaU}

Supersymmetry, as most extensions of the SM, introduces new sources of FCNC
and/or of
CP violation. In general, this simply happens because of the extended
particle spectrum,
which allows new flavour changing and/or CP violating interactions to be
written down.

\begin{table}
\begin{center}
\begin{tabular}{|l|l|l|l|}\hline
\parbox{3cm}{\centering{\bf Universality} $m^2\propto 1$}&
\parbox{3cm}{
\hbox{(a) Min SuGra}
\hbox{(b) Gauge Mediation}
\hbox{(c) Fixed Points}} &
\parbox{1cm}{No\\ Yes\\ ?} &
\parbox{3cm}{$V_{\rm CKM}$-like ($b\to s \gamma$, $b\to s \ell^+\ell^-$)}\\
\hline
\parbox{3cm}{\centering\bf Alignment} &
\parbox{3cm}{$\lambda_{\rm d}$, $m^2_{\tilde{\rm d}}$\\ simultaneously\\ diagonal} &
No (?) & $d_N$, CP($D$)\\ \hline
\parbox{3cm}{\centering\bf Heavy $\tilde{f}_{1,2}$} & 
\parbox{3cm}{$\varepsilon_K$ adjusted}&
Yes  &
\parbox{3cm}{$d_N, d_e, \mu\to e\gamma,\varepsilon_K$, CP($B$), $\varepsilon'/\varepsilon$}\\
\hline
\parbox{3cm}{\centering\bf Non abelian flavour symmetry} &
\parbox{3cm}{$\tilde{m}_1\approx \tilde{m}_{2,3}$\\$W\approx V_{\rm CKM}$}&
Yes  &
\parbox{3cm}{$d_N, d_e, \mu\to e\gamma,\varepsilon_K$, CP($B$)}\\
\hline
\end{tabular}
\caption{\em Supersymmetry and FCNC/CP. The third column comments on
the compatibility with Unification. The last column indicates the observables
which are likely to be affected.}
\end{center}
\end{table}

Some of these new interactions are still controlled by the same Cabibbo
Kobayashi Maskawa
matrix,
$V_{CKM}$, which appears in the standard charged current weak interactions.
The related new
effects are  there no matter how supersymmetry is broken and are therefore
present in any
realistic supersymmetric extension of the SM. Observables that could be
 affected in a significant way
by these new sources of flavour violations, depending on the spectrum
of the superpartners, are the inclusive $b
\rightarrow s(d) +
\gamma$~\cite{bsg} and  $b \rightarrow s(d) + l^+ l^-$~\cite{bsl} rates.
The current
experimental average  branching ratio for
$b \rightarrow s + \gamma$, $(2.55 \pm 0.61) 10^{-4}$~\cite{drell},
compared with the pure SM
prediction,
$(3.48 \pm 0.31) 10^{-4}$~\cite{bsgt}, still leaves room for a sizeable new
contribution,
negatively interfering with the SM amplitude.

In general, however, other sources of FCNC/CP-violations are present, not
controlled
by the CKM matrix. A necessary condition for their physical relevance  is the
non degeneracy, in flavour space, of the supersymmetry breaking masses of
the scalar partners
of fermions with given charge and chirality. If supersymmetry and flavour
breaking are
decoupled, as in "gauge-mediation-type" models, no such extra source of
FCNC/CP-violation
needs to be present. In the opposite case, like in "supergravity-type"
models, they are
present~\cite{nw} and expected to be sizeable if Grand Unification is
realized~\cite{hkr,bh}.
In general these effects are described by new unitary mixing matrices
occurring in the gaugino
vertices, called $W$ hereafter, and by some other FCNC/CP-violating
parameters in the analytic scalar potential. Generically, these models are
actually in
serious danger of conflicting with existing observations, or non
observations, due to
supersymmetric FCNC/CP-violating loop effects~\cite{ant}.

For this
reason, several ideas have been recently put forward to avoid this problem,
referred
to as the "supersymmetric flavour problem". An attempt to summarize them is
made in Table 2,
also including the "Universality" option. Given the variety of
possibilities, there is a significant amount of arbitrariness in preparing
this Table. To
balance this arbitrariness, it will be useful to consider the original
references and also
the discussion given by Nir~\cite{nir} at this Conference on related matters.

I have already commented upon the "Universality" case~\cite{dg1}, which can
certainly be
realized in  "gauge-mediation-type" models~\cite{gmn}. It is possible to
get close enough to
it even in supergravity, with universal boundary conditions, if no
Unification is realized,
since the flavour-breaking renormalization effects due to the top Yukawa
couplings,
$\lambda_t$,~\cite{nw} are small enough and, most importantly, they do not
introduce new
CP-violating phases nor they affect leptons. Maybe a dynamical way of
realizing "Universality"
in supergravity exists, using strong renormalization effects which could
lead the
supersymmetry breaking masses to flavour-universal fixed points~\cite{su}.
However, I have
not yet seen nor have been able to find myself a fully realistic
realization of this idea.

The ideas that try to solve the "supersymmetric flavour problem" without
resorting to flavour-degenerate scalar masses can perhaps be grouped into
three distinct
categories: i) Alignment; ii) Heavy $\tilde{f}_{1,2}$; iii) Non Abelian Flavour
Symmetries.

In its
simplest version, Alignment~\cite{alig} is the notion that some Abelian
flavour symmetry may
force the Yukawa coupling matrix responsible for the down-type quark to be
(approximately)
diagonal with the down s-quark mass matrices in
the same superfield basis. This is as saying that  the neutral gaugino
interactions of the
down quark-squarks will not involve any non trivial
$W$-mixing matrix, thus keeping under control an otherwise very problematic
effect in
$\Delta S=2$ transitions, especially related to the $\epsilon$ parameter in
kaon
physics~\cite{ant}. One should not forget here that another very stringent
constraint on
FCNC/CP effects comes from the non observation, so far, of any Lepton
Flavour Violating
process like $\mu
\rightarrow e + \gamma$~\cite{ant}. To satisfy this new constraint requires
an alignment in
the lepton sector too. Although this is possible, I do not know how to make
all this
consistent with Unification. The most likely signatures of Alignment are in
any case related
to the mixing matrices $W$ in the up quark-squark sector, i.e. occurring in
the neutron
Electric Dipole Moment, via the u-quark EDM, and in mixing or CP-violation
in the charm
system.

The fact that the most dangerous FCNC/CP effects come from exchanges of non
degenerate
scalars of the first and second generations can be put togheter~\cite{ckn},
in an attempt to
alleviate the problem, with the observation, mentioned in Sect.1, that
these same scalars
can be considerably heavier than those ones of the third generation,
between 1 and 10 TeV,
without conflicting with the naturalness constraint~\cite{dgi}. This
possibility is called
``Heavy
$\tilde{f}_{1,2}$'' in Table 2. Although this is enough to suppress the LFV
processes, this is not
the case for the
$\epsilon$ parameter, which requires a further suppression by about two
orders of magnitude,
maybe coming from the smallness of the relevant phase(s).

Finally, is it not possible that the same flavour symmetry that keeps the CKM
matrix and the $W$-matrices close to $\bf 1$, also forces the first and
second generation
scalars to be degenerate enough to solve the $\epsilon$ and the $\mu
\rightarrow e + \gamma$ problems automatically? Among many different
alternatives~\cite{nafs},
a clear candidate emerges for such a symmetry, which is consistent with
Unification: a $U(2)$
symmetry acting on the three generations of matter fields $ \Psi_i,
i=1,2,3$, as a doublet
plus a singlet,
\begin{equation}
\Psi_i = \Psi_{\alpha} + \Psi_3 , \alpha = 1,2,
\label{psi}
\end{equation}
and trivially on the Higgs fields~\cite{pt,u21,u22}.
Whereas U(3) is the largest flavour symmetry group in
presence of full vertical Unification and for vanishing Yukawa couplings,
U(2) is the leftover
subgroup after taking into account the large $\lambda_t$. The key observation
is that, in the limit of unbroken U(2), the first two generations of
fermions are massless,
$m_{f1} =m_{f2} = 0$, the first two generations of sfermions are
degenerate, $\tilde{m}_{f1}
=\tilde{m}_{f2} $ and the CKM-matrix is $\bf
1$, as are the
$W$-matrices.  Therefore, a simple pattern of small U(2)-breaking can
correlate the small fermion masses $m_{f1}$ and
$m_{f2}$ to slightly non degenerate sfermions of the first two generations
, $\tilde{m}_{f1}
\not=\tilde{m}_{f2} $, and to small deviations from unity of both $V_{CKM}$ and the
$W$-matrices~\cite{u22}.

The interesting outcome of all this, other than the correlation between
fermion masses and the CKM parameters, is that the significant FCNC/CP
 effects from
supersymmetric loops come from the third generation of sfermions not being
degenerate with
the first two and with $W$-mixing angles similar to the corresponding CKM
angles. In turn,
this gives effects of naturally similar magnitude as the SM effects both in
$\epsilon$ and in
$B-\bar{B}$ mixing, which all come from top exchanges. This is consistent with
present observations. It should no longer be, however, after the
experiments foreseen in
B-physics in the next few years~\cite{nir,u22,bphys}.

The U(2) symmetry does not give any control on the splitting
among the third and the first or second generations of scalars. However, as
mentioned
previously, already the renormalization effects due to the large
$\lambda_t$ in a Unified Theory induce splittings of relative order unity
both in the
s-quark and in the s-lepton case~\cite{bh,bhs12}. The splitting among s-leptons is
also particularly interesting: it is a fact that third generation s-leptons
significantly non
degenerate with the first two generations and with
$W$-mixing angles similar to the corresponding CKM angles give LVF effects,
in $\mu
\rightarrow e + \gamma$ or in $\mu
\rightarrow e $ convertion in atoms, close to the current limits for
sparticle masses in their
full natural range, as defined above. The same remark applies to the electron
EDM~\cite{dh,bhs12}, if the new phase in the relevant $W$-matrix is of
order unity.

\section{Conclusions} \label{sec: concl}

The theoretical case for Unified Theories is strong and supported by a piece
of empirical evidence, the successful prediction of the strong coupling
constant.

Means
exist to establish or strongly  indicate supersymmetric Unification
experimentally, some of
which  compelling: the searches for supersymmetric particles and for a
light Higgs, both
within reach of facilities now active or under construction. Motivated by
supersymmetry and
Unification are also the searches for proton decay and for dark matter in
the form of WIMPs.

Finally,
several observables in flavour physics are likely to be affected in a
significant way by
virtual exchanges of supersymmetric particles. This will have to be the
case, I believe, if
 Unification is realized in the supergravity way, with flavour breaking
coupled to supersymmetry breaking. In some cases, like the
$\epsilon$ parameter and mixing or CP violation in the B-system, the
supersymmetric effects
compete with the SM effects, from which they have to be disentangled. In
other cases, like
 in $\mu
\rightarrow e + \gamma$ and in the electron and neutron EDMs, these effects
stand as
unambiguous signals of new physics. Their finding, togheter with the study
of the CKM
parameters, might also shed light on the role, so far elusive, of flavour
symmetries. I consider the experimental effort in these directions very worthful.

\section*{Acknowledgements}

I wish to thank for many useful discussions and comments G. Altarelli, 
G. Bhattachariya, S. Dimopoulos, G. Dvali, G. Giudice, L. Hall, 
R. Rattazzi, S. Raby, G. Ross, A. Romanino and 
A. Strumia. I am indebted to A. Strumia for the help in the
editing of these pages.

\end{document}
\\
Title: Unified theories
Authors: R. Barbieri
Comments: 11 pages, 2 figures.
Report-no: talk given at LP97
\\
The present status of Unified Theories is summarized with special emphasis
on their possible experimental tests. Outline: i) Unification
of couplings; ii) Where can a positive signal come from? iii) HERA anomaly
and Unification; iv) Recent progress in model building; v) Flavour and
Unification.
\\